\documentclass[amssymb,twocolumn,showpacs,prl,aps]{revtex4}
\usepackage{graphicx}
\usepackage{amssymb}
\begin{document}

\title{
       Comment on "Collective dynamics in liquid lithium, sodium, and aluminum"
      }
\author{
        Tullio Scopigno$^{1}$ and Giancarlo Ruocco$^{1}$
        }
\affiliation{
    $^{1}$Dipartimento di Fisica and INFM, Universit\'a di Roma ``La Sapienza'',
I-00185, Roma, Italy.\\
    }
\date{\today}

\begin{abstract}

In a recent paper, S. Singh and K. Tankeshwar (ST), [Phys. Rev. E
\textbf{67}, 012201 (2003)], proposed a new interpretation of the
collective dynamics in liquid metals, and, in particular, of the
relaxation mechanisms ruling the density fluctuations propagation.
At variance with both the predictions of the current literature
and the results of recent Inelastic X-ray Scattering (IXS)
experiments, ST associate the quasielastic component of the
$S(Q,\omega)$ to the thermal relaxation, as it holds in an
ordinary adiabatic hydrodynamics valid for non-conductive liquids
and in the $Q \rightarrow 0$ limit. We show here that this
interpretation leads to a non-physical behaviour of different
thermodynamic and transport parameters.

\end{abstract}


\maketitle


Recently, Singh and Tankeshwar (ST) reported a new version of the
finite wavevector, $Q$, extension of the generalized hydrodynamics
(the so called "molecular hydrodynamics") applied to the dynamics
of simple liquid metals \cite{sing_pre}. The goal of the ST paper
is twofold. On one side, the authors propose an analytic
expression for the density-density correlation function, $F(Q,t)$,
which is based on the existence of two relaxation channels, and
they test it through its Fourier transform against the dynamic
structure factor recently measured by inelastic X-ray scattering
(IXS) on several liquid metals. On the other side, the authors
assign a specific meaning to these two relaxation processes: the
slower process is traced back to the thermal relaxation, while the
faster one is ascribed to viscous effects. More specifically, at
variance with the results of previous studies performed on the
same IXS data used by ST \cite{scop_prlli,scop_preal,scop_prena},
they assign the quasielastic component of the IXS spectra to the
thermal relaxation.

The aim of this comment is to point out how, although the
analytical spectral shape proposed by the authors turns out to
reproduce the experimental spectra at a reasonable level of
accuracy, the interpretation behind it is inadequate to describe
the microdynamics of simple metals. In particular, assigning the
whole quasielastic IXS spectrum to the thermal relaxation, the
value and the $Q$-dependence of the derived fitting parameters
are not consistent with independent determination of the transport
(viscosity, thermal conductivity) and thermodynamic (specific
heats) properties and disagrees with what it is expected on a
general ground.

The dynamic structure factor of simple liquids, in the $Q
\rightarrow 0$ limit, is constituted by three lines, a
quasielastic, lorenzian, component centered at zero frequency, and
a doublet (Brillouin peaks) symmetrically shifted at finite energy
\cite{BERNE}. In this limit, the energy position of the side
lines, their broadening and the broadening of the central line are
successfully described within the well known simple hydrodynamics
approach. In particular, the frequency of the Brillouin mode is
$\omega_B=c_0Q$, being $c_0$ the adiabatic sound velocity, while
its linewidth (HWHM) is ruled by thermal ($\Gamma_{th}$) and
viscous ($\Gamma_{\eta}$) damping as

\begin{equation}
\left \{
\begin{array}{ll}

\Gamma&=\Gamma_{\eta} + \Gamma_{th}\\

\Gamma_{\eta}&=\frac{Q^2}{2\rho}\left[\frac{4}{3}\eta_s+\eta_B
\right]\\

\Gamma_{th}&=\frac{Q^2}{2\rho}\left[(\gamma -1)\lambda/C_p \right]

\label{gidro}
\end{array}
\right.
\end{equation}

\noindent where $\lambda$, $\gamma$ and $C_p$ are the thermal
conductivity, specific heat ratio and constant pressure heat
capacity, while $\eta_s$ and $\eta_B$ are the shear and bulk
viscosity, respectively. The width of the quasielastic mode,
instead, is $1/\tau_{th}=D_TQ^2$, i.e. it is ruled by the thermal
diffusion only.

In Ref.\cite{sing_pre}, ST extend this approach to the finite
wavevector domain probed by IXS, and they apply it to a specific
sub-class of simple liquids, namely liquid metals. To this
purpose, ST replace the lorenzian shape stemming from simple
hydrodynamics with a hyperbolic secant shape, leaving unchanged
the origin of the individual contributions.

The extension to finite wavevectors, however, in the special case
of highly conductive systems, requires careful evaluation of the
physics behind the model, for two main reasons. First, as pointed
out in the celebrated textbook by Faber \cite{FABER}, in a liquid
metal, owing to the high thermal conductivity, on pushing the
wavevector to values comparable to the inverse mean interparticle
distance (as in the present case) the quantity $D_TQ^2$ soon
becomes larger than the Brillouin frequency $\omega_B$.
Consequently, as predicted (ref. \cite{FABER}), the thermal peak
broadens ultimately overlapping with the Brillouin lines, the
sound propagation turns from adiabatic to isothermal, and
independent thermal fluctuations become impossible. A second
reason is that, on the snapshot timescale probed by IXS (Thz), the
diffusive atomic motion caracteristic of the liquid state looks
frozen, and therefore part of the viscous contribution ruling the
Brillouin width at low $Q$ is transferred to the elastic line.
Both of these reasons have been disregarded by ST.

A formal way to quantify the scenario qualitatively depicted
above, is to introduce the memory function formalism. Within this
framework, the normalized density autocorrelation function
$\phi(Q,t)=F(Q,t)/S(Q)$ obeys the Langevin equation

\begin{eqnarray}
\stackrel{..}{\phi}(Q,t)+\omega_0^2(Q)\phi(Q,t) +\int_0^tM(Q,t-t^{\prime })%
\stackrel{.}{\phi}(Q,t^{\prime })dt^{\prime }=0  \label{langevin}
\end{eqnarray}

\noindent where $\omega_0^2(Q)=-\stackrel{..}{\phi}(Q,0)=K_B T
Q^2/ m S(Q)$ is related to the generalized isothermal sound
velocity $c_t (Q)=\omega_0 (Q) / Q$ and $M(Q,t)$ is the second
order memory function which, in the hydrodynamic limit reads:

\begin{equation}
M(Q,t) = 2D_VQ^2\delta (t) + \left( \gamma -1\right) \omega
_{0}^{2}(Q)e^{-D_{T} Q^{2}t} \label{memory}
\end{equation}

\noindent with $D_V=(\frac{4}{3}\eta_s+\eta_B) / \rho$ the
longitudinal viscosity and $\gamma$ the specific heat ratio. It
is easy to show from Eq. (\ref{langevin}) and (\ref{memory}),
indeed, that \textit{if} the condition $\omega_B>>D_TQ^2$ holds,
the dynamic structure factor (i.e. the Fourier trasform of $F$)
reduces to the sum of a (thermal) central lorentian contribution
and a damped harmonic oscillator spectrum of characteristic
frequency $\omega_B \approx \sqrt{\gamma} \omega_0$ (which is the
sum of two symmetrically shifted lorenzians in the limit
$\Gamma<<\omega_0$). As $\gamma=(c_0/c_t)^2$, one ultimately gets
the hydrodynamic result, i.e. an adiabatic regime.

The finite $Q$ wavevector generalization stems dropping the
hypothesis of the instantaneous (Markovian) nature of the viscous
term. In particular, one can introduce a finite timescale ($\tau$)
for the decay of the first term in Eq. (\ref{memory}) and,
according to the $\omega_0\tau $ value, the viscous relaxation
can affect both the quasielastic and the inelastic peaks.
Actually, the situation is even more involved. In earlier MD
studies of liquid metals it has been soon realized that the
viscous dynamics in the microscopic regime (i.e. at wavelength
comparable with the inverse mean inter-particle separation)
proceeds through two distinct processes, characterized by two
well separate time scales ($\tau_\alpha$ and $\tau_\mu$)
\cite{BALUCANI}. This idea has been recently substantiated by a
number of IXS investigation, where the presence and the role of
the two relaxation mechanisms has been identified and widely
discussed
\cite{scop_jpc,scop_prlli,scop_preal,scop_prena,scop_prlga}. In
particular, it has been pointed out that the slower (-$\alpha$)
relaxation time satisfies the condition $\omega_B(Q)
\tau_\alpha(Q)>>1$, i.e. some part of the viscous flow is frozen.
As a consequence, at the wavevectors typical of the IXS
experiments ($Q=1\div 20$ nm$^{-1}$) the quasielastic spectrum
acquires a component arising from this frozen structural
relaxation. Moreover, owing to the high thermal conductivity, in
liquid metals the condition $\omega_B \tau_{th} <1$ holds down to
wavevectors of the order of 0.1 nm$^{-1}$ \footnote{This estimate
is valid in the case of interest, i.e. in a wavevector region not
to close to the main peak of the structure factor, where
structural effects are expected to occur, so that one can
reasonably assume $\omega(Q) \approx cQ$ and $D_T(Q)\approx
D_T(Q\rightarrow 0)$}. In this limit, the sound propagation is
isothermal and not adiabatic, which means that the thermal
relaxation is too broad to give a noticeable quasielastic
component, while it renormalizes the sound velocity from the
hydrodynamic value $c_0$ to the isothermal value $c_t$, as
clearly discussed in Ref. \cite{scop_jpc} and pointed out in
previous literature \footnote{For example, in ref. \cite{FABER}
the discrepancy observed in liquid lead between the sound
velocity measured with ultrasound (adiabatic) and with inelastic
neutron scattering has been tentatively assigned to the
isothermal nature of the sound propagation at the probed
wavevectors.}.

Summing up, the quasielastic peak in the $S(Q,\omega)$ is
expected to arise from the frozen structural relaxation and not
from the thermal process. Its linewidth is therefore associated
to $\tau_\alpha ^{-1}$ and not to $D_TQ^2$.

A decisive support to the origin that we propose for the
quasielastic peak can be gained by the wavevector and temperature
dependence of the spectrum. The signature of a viscous origin of
the central peak are, indeed, i) a substantial $Q$-independence
($\tau_{\alpha}$ is almost $Q$ independent), and ii) a strong
temperature dependence of its width (the structural relaxation
follows the Arrhenius or an even faster behaviour \cite{ang2}). On
the contrary, a central peak of thermal origin must show i) a
strong $Q$ dependence (width $\propto Q^{2}$ until the structure
factor is flat) and ii) a very weak temperature dependence (traced
back to the mild temperature dependence of the thermal
conductivity). While the temperature dependence of the central
peak is difficult to determine experimentally, because of the
impossibility to supercool liquid metals (some hints could came
from the molecular dynamics simulation, see below), the $Q$
dependence of the central peak is rather clear: its experimental
width (as determined by IXS spectra) is a quantity that can be
roughly determined by a ruler, or better by a fitting procedure,
and turns out to be almost constant in the examined $Q$-range. If
one interpreted (following ST) this width as due to the thermal
relaxation process, it would be equal to $D_T Q^2$ and, as a
consequence, the generalized thermal diffusivity, would obviously
become strongly $Q$-dependent ( $D_T(Q) \propto Q^{-2}$ ),
ultimately dropping by a factor $\approx 400$ from the
hydrodynamic limit to the examined $Q$ range. This behaviour is
difficult to understand on a physical ground, and is very
different from that observed in other liquids. Indeed, although we
are not aware of any explicit $D_T(Q)$ estimates for liquid
metals, detailed calculations for other liquids (water
\cite{bert_water}) show a decrease of only a factor ten on going
from $Q$=0 and the $Q$ value of the the maximum of the static
structure factor. The too strong $Q$ dependence of $D_T(Q)$
derived by ST speaks against the assignment of the central peak to
the thermal process. Turning now our attention to the temperature
dependence of the central peak width, a recent MD work on
underccoled liquid lithium showed that this width is actually
strongly temperature dependent. More important, the derived
relaxation time, $\tau_\alpha$, closely follows the behaviour of
the mass diffusion coefficient (as expected, for example, from the
predictions of the Mode Coupling Theory for the structural
relaxation time \cite{gotz_jpc}), increasing by more than a factor
ten in the spanned temperature range \cite{scop_presim}. Such a
strong temperature dependence is not expected for the thermal
diffusivity, indicating once more the non-thermal origin of the
central peak.

\begin{figure} [t]
\centering
\includegraphics[width=.5\textwidth]{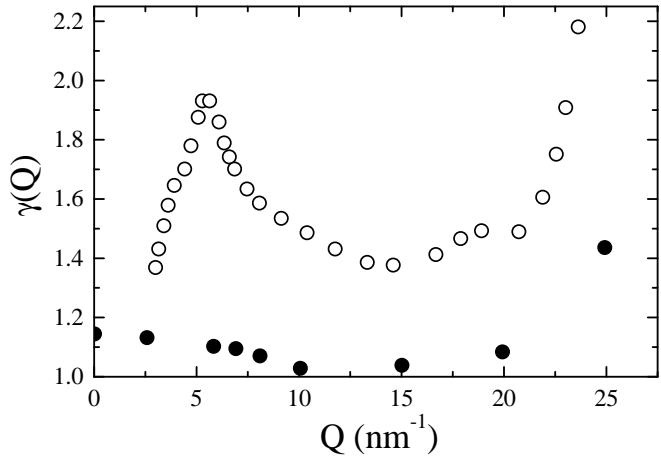}
\vspace{-7cm} \caption{Comparison between the generalized specific
heat ratio $\gamma=c_p/c_v$ as obtained by the ST model ($\circ$)
and by direct numerical calculation ($\bullet$)} \label{fgamma}
\end{figure}

A further consequence of the assignment of the central peak to the
thermal relaxation process is the anomalous $Q$ dependence,
obtained by ST, of the parameter describing the ratio of the
elastic to inelastic scattering intensity. In the authors
notation this is the quantity $a(Q)$ entering in their Eq. (4),
which is related to the generalized specific heat ratio $\gamma
(Q)$ via the expression $\gamma=\frac{1}{1-a}$. In
Fig.\ref{fgamma} we report $\gamma (Q)$ as obtained by ST
compared to previous numerical calculation for the case of liquid
lithium \cite{can_lit}. The values from the ST analysis clearly
overestimates the calculated ones, although these latter have
been obtained through simulations which well reproduce the
experimental spectra.

Finally, the hydrodynamic expression (Eq. (\ref{gidro})) which ST
utilize to estimate the sound attenuation (and to judge the
reliability of their model) is valid only for ordinary (non
conducting) liquids in the $Q \rightarrow 0$ limit. It accounts,
indeed, for the first-order non vanishing attenuation
contribution from the thermal diffusive mode, and it relates the
attenuation to the whole longitudinal viscosity. As pointed out
in Ref. \cite{scop_jpc}, the small attenuation contribution from
the thermal process in the IXS wavevectors range has, in liquid
metals, the different expression

\begin{equation}
\Gamma_{th}=\left( \gamma -1\right) c_t^2/D_T
\end{equation}

\noindent being $c_t$ the isothermal sound speed. More important,
at the probed wavevectors, only the microscopic, relaxed, part of
the viscosity is involved in the sound damping $\Gamma_\eta$, as
the structural relaxation contribution if frozen on the probed
time-scale (in liquid lithium this latter contribution from the
microscopic part is one half of the whole viscous term
\cite{scop_jpc}).

In conclusion, in our opinion the model proposed by Singh and
Tankeswar has the merit to indicate a possible heuristic
analytical time- (and frequency-) dependence of the intermediate
scattering function (dynamic structure factor) appropriate to
catch some features of the dynamics in the $Thz$ frequency region.
Indeed, for instance, the functional form proposed by ST has all
the spectral moment finite, while, on the contrary, the memory
function approach, leading to spectral shape that are described by
fraction of polynoms, has only a finite number of converging
spectral moments. The ST analysis, however, seems to be rather
weak in the physical interpretation of the relaxation processes
that drive the collective dynamics. None of the two parameters
$\tau_1$ and $\tau_2$ has a clear meaning and can be directly
associated to any relaxation mechanism, as it happens for example
within a memory function framework. As a consequence of the
incorrect identification of the role of the thermal process, the
parameter $\tau_1$ has no connection with any thermal property
(being in fact mainly related to the structural relaxation time
$\tau_\alpha$), the generalized specific heat ratio obtained from
the fitting shows marked discrepancies with the expected
behavior, and the viscosity can not be estimated through Eq.~(11)
in Ref.~[1].

We thank D. Fioretto, J.P. Hansen, F. Sacchetti and G. Senatore
for stimulating discussions. Fruitful email exchange with S. Singh
and K. Tankeshwar is also acknowledged.


\end{document}